\documentclass[fleqn,usenatbib]{mnras}

\usepackage{newtxtext,newtxmath}

\usepackage[T1]{fontenc}

\DeclareRobustCommand{\VAN}[3]{#2}
\let\VANthebibliography\thebibliography
\def\thebibliography{\DeclareRobustCommand{\VAN}[3]{##3}\VANthebibliography}

\usepackage[T1]{fontenc}
\usepackage[utf8x]{inputenc}
\usepackage[english]{babel}
\usepackage{graphicx}
\usepackage{dcolumn}
\usepackage{bm}
\usepackage{physics}
\usepackage{here}
\usepackage{mathrsfs}
\usepackage{color}
\usepackage{amsmath}

\title[Impacts of hydrogen envelope on supernova fallback]{Impacts of hydrogen envelope on supernova fallback and the resulting compact remnant masses}

\author[Shinoda et al.]{
Kengo Shinoda$^{1,2},$\thanks{E-mail: kengo0317@g.ecc.u-tokyo.ac.jp}
Yudai Suwa$^{1,3}$,
Ryosuke Hirai$^{4,5,6}$,
Ryo Sawada$^{7}$,
Kengo Tomida$^{8}$,
Kazunari Iwasaki$^{9}$,
\newauthor
Takeru K. Suzuki$^{1,2}$
\\
$^{1}$Department of Earth Science and Astronomy, The University of Tokyo, Tokyo 153-8902, Japan\\
$^{2}$Department of Astronomy, The University of Tokyo, Tokyo 113-0033, Japan\\
$^{3}$Center for Gravitational Physics and Quantum Information, Yukawa Institute for Theoretical Physics, Kyoto University, Kyoto 606-8502, Japan\\
$^{4}$Astrophysical Big Bang Laboratory, Pioneering Research Institute, RIKEN, 2-1 Hirosawa, Wako, Saitama 351-0198, Japan \\
$^{5}$School of Physics and Astronomy, Monash University, Clayton, Victoria 3800, Australia\\
$^{6}$OzGrav: The ARC Centre of Excellence for Gravitational Wave Discovery, Australia\\
$^{7}$Institute for Cosmic Ray Research, The University of Tokyo, Kashiwa, Chiba 277-8582, Japan\\
$^{8}$Astronomical Institute, Tohoku University, Sendai, Miyagi, 980-8578, Japan\\
$^{9}$Center for Computational Astrophysics, National Astronomical Observatory of Japan, 2-21-1 Osawa, Mitaka, Tokyo 181-8588, Japan
}

\date{Accepted XXX. Received YYY; in original form ZZZ}

\pubyear{\the\year{}}

\begin{document}
\label{firstpage}
\pagerange{\pageref{firstpage}--\pageref{lastpage}}
\maketitle

\begin{abstract}
Fallback in core-collapse supernovae plays a central role in setting compact-remnant masses and may produce late-time emission. In hydrogen rich progenitors, the reverse shock arising at the hydrogen-helium interface has the potential to dramatically enhance fallback, yet its overall impact across a broad explosion-energy range has not been systematically quantified.
Using one-dimensional hydrodynamic simulations for metal-poor progenitors with $M_{\rm ZAMS}=18$--$28\,M_\odot$ and models with and without hydrogen envelopes, we explore fallback over explosion energies of $10^{48}$--$10^{52}\,{\rm erg}$. We find a robust and universal mass-transition behaviour: when the explosion energy reaches only $2$--$3$ times the binding energy of the hydrogen envelope, the reverse shock returns to the centre and sharply increases the remnant mass by $\gtrsim 2\,M_\odot$. Above this threshold, the reverse shock escapes and hydrogen-rich and stripped-envelope progenitors yield nearly identical remnant masses.
By normalizing the results with the envelope binding energy, we show that all progenitor models converge to a common fallback relation. We further provide a simple analytic prescription that connects explosion energy, hydrogen-envelope binding energy, and final compact-remnant mass. This relation provides an important link between progenitor properties and compact-remnant masses, and is useful for population-synthesis and galactic chemical-evolution studies.
\end{abstract}




\section{Introduction}
\label{sec:Introduction}
Core collapse supernovae (CCSNe) are explosions that occur during the final stages of stellar evolution in massive stars with masses exceeding $8\, \mathrm{M_\odot}$. 
During the explosion, a neutron star or black hole forms at the centre, leading to a phenomenon known as fallback accretion \citep{Colgate1971ApJ, Chevalier1989ApJ, Woosley1995ApJS, MacFadyen2001ApJ, Zhang2008ApJ}. 
In this process, a part of the ejecta falls back onto the central compact object due to its gravitational field. 
Fallback accretion increases the mass and spin \citep{Janka2013MNRAS, Wongwathanarat2013A&A,Chan2020MNRAS,Janka2022ApJJ} and changes the magnetosphere \citep{Muslimov1995ApJ,Geppert1999A&A,Bernal2010RMxAA,Bernal2013ApJ,Shigeyama2018PASJ, Zhong2021ApJ, Inoue2025arXiv} of the compact object. Therefore it is important for interpreting the observed variety of compact objects \citep[e.g., black hole mass gap;][]{Bailyn1998ApJ, Ozel2010ApJ, Farr2011ApJ, Ugliano2012ApJ, Abbott2023PhRvX}.
On the other hand, recent studies have also proposed that supernova fallback can contribute to the formation of long-period radio pulsars \citep{Tan2018ApJ,Han2021RAA,Caleb2022NatAs,Hurley-Walker2022Natur}. If the fallback material forms a disc around the neutron star, its interaction with the magnetosphere of the neutron star may extract angular momentum and lead to a substantial spin-down. This mechanism allows spin periods of $\sim 100\,{\rm s}$ or longer \citep{Gencali2022MNRAS,Ronchi2022ApJ,Gencali2023MNRAS}.
Moreover, several studies suggest that fallback not only affects neutrino emission \citep{Akaho2024ApJ, Nakazato2024ApJ}, but may also be enhanced by neutrino cooling, which triggers accretion exceeding the Eddington limit \citep{Houck1991ApJ,Zampieri1998ApJ,Perna2014ApJ}.

However, many aspects of fallback physics remain unsolved, and only a few studies have focused on the relationship between the fallback and the progenitor's hydrogen envelope. 
If the progenitor has a hydrogen envelope, the interaction of shock waves at the hydrogen-helium core interface can lead to the formation of both forward and reverse shocks \citep{Hachisu1990ApJ,Zhang2008ApJ, Dexter2013ApJ,Vartanyan2025ApJ}. 
Although more than $50$ percent of CCSNe originate from progenitors with hydrogen envelopes \citep{Li2011MNRAS, Ma2025arXiv}, it remains unclear whether the reverse shocks fall back onto the central compact object.

\citet{Zhang2008ApJ} showed that in zero-metallicity 25 $M_\odot$ hydrogen-rich progenitors with an explosion energy of $E_\mathrm{exp}= 1.2 \times 10^{51}\, {\rm erg}$, the reverse shock increases the remnant mass by approximately 1.7 $\mathrm{M_\odot}$. In contrast, in solar-metallicity hydrogen-poor progenitors, the reverse shock remained above the escape velocity, preventing fallback to the compact object. 
These results indicate that the hydrogen envelope has a strong impact on fallback via the reverse shock.
However, it is uncertain how much of this effect arises purely from the envelope mass,
since the two models differ in their internal core structures.

Observationally, both hydrogen-rich and hydrogen-poor supernovae are found, reflecting the diversity in the degree of envelope stripping among progenitors.
The Carnegie Supernova Project has shown that type II SNe, which retain a hydrogen envelope, exhibit a wide range of explosion energies \citep{Martinez2022A&A}.
However, no previous study has systematically isolated the effect of the hydrogen envelope itself across a wide range of explosion energies.
Most existing works have compared progenitor models that differ not only in envelope mass but also in core structure, making it difficult to assess the envelope’s independent impact on fallback.

Population synthesis studies of systems involving compact remnants (BHs and NSs) often use prescriptions that assume a one-to-one mapping between the C/O core mass of the supernova progenitor and the final remnant mass \citep{Fryer2012ApJ, Mandel2020MNRAS}. However, if there is substantial fallback, the remnant mass will also depend on the hydrogen envelope mass and therefore this one-to-one relation with the C/O core mass will not hold. To provide more accurate predictions for the mapping between SN progenitor properties and their remnant masses, it is essential to isolate the contribution of the hydrogen envelope itself to the fallback process and to quantify its effect across a wide range of explosion energies.

In this paper, we aim to clarify the correlation between the structure of the hydrogen envelope and the fallback mass across a wide range of explosion energies, from failed SNe \citep{Nadezhin1980Ap&SS, Fernandez2018MNRAS, De2024arXiv} to canonical SNe. The structure of the paper is as follows. In Section \ref{sec:method}, we describe our simulation method, setup, and the construction of initial conditions for hydrodynamical simulations. In Section \ref{sec:results}, we analyse the shock propagation in CCSNe, its fallback accretion rate, and the time evolution of central compact object mass, comparing progenitors with and without hydrogen envelope. In Section \ref{sec:discussion_and_conclusion}, we discuss our results and present conclusions.

\section{Method}
\label{sec:method}
 In this study, we perform a suite of supernova explosion simulations to investigate the amount of fallback and how it depends on various physical parameters. We solve the hydrodynamics equations without magnetic fields, including the self-gravity and cooling terms. Our simulations are performed with the public code {\tt Athena++} \citep{ Stone2020ApJS} in one-dimensional spherical symmetry.

We make two structural modifications to the progenitor of \cite{Woosley2002RvMP}.
The first is a modification to the hydrogen envelope. 
In Section \ref{subchap:envelope_method},  we show how we artificially generate profiles of progenitors without hydrogen envelope (SESN) that have the same core properties as a given progenitors with hydrogen envelope (SNII).
The second is a change to the core structure. 
In Section  \ref{subchap:softened_method}, we explain the core softening method \citep{Ohlmann2017A&A,Hirai2020MNRAS}, which remodels the steep density structure of the core to flatter profile.

\subsection{Equations}
The basic equations are written as follows, 
\begin{align}
\frac{\partial \rho}{\partial t} + \nabla \cdot (\rho \mathbf{u}) &= 0, \\
\pdv{\rho \mathbf{u}}{t} + \nabla \cdot (\rho \mathbf{u}\otimes \mathbf{u}+p\mathbf{I}) &= \rho \mathbf{g}, \\
\pdv{e}{t} + \nabla \cdot \left\{\mathbf{u} (e + p) \right \} &= \rho \mathbf{u} \cdot \mathbf{g} - Q_\mathrm{cool},
\end{align}
where $\rho$ is the density, $\rho\mathbf{u}$ is the momentum density, and $e = \varepsilon + \rho \mathbf{u}\cdot\mathbf{u} / 2$ is the sum of kinetic energy density and internal energy density. 
$p$ is the gas pressure, and $\mathbf{I}$ is the identity tensor.
The adiabatic index of the equation of state  (EoS) is set to $\gamma = 5/3$ 
\footnote{In our calculation domain, the ratio between radiation pressure and total pressure remains around $\sim 0.3$. For simplicity, we therefore adopt an adiabatic index of $\gamma = 5/3$. The effects of including radiation pressure, matter--radiation coupling, and ionization are discussed in detail in \citet{Hirai2020MNRAS}.} 
and the internal energy density is given by $\varepsilon = p/(\gamma -1)$. One of the source terms is self-gravity, expressed as $\mathbf{g}=-\nabla\phi$, where $\phi(r) = -\int^\infty_r dr' GM_{r'} / {r'}^2 (r \geq r_{\rm s})$ is the gravitational potential outside the region where the core softening method (see Section \ref{subchap:softened_method}) is applied. 
Here, $M_r$ denotes the enclosed mass within radius $r$. The cooling term $Q_\mathrm{cool}$ and softening radius $r_{\rm s}$ defined in Section \ref{subchap:softened_method}.

\subsection{Numerical Setup}
We use the public code {\tt Athena++} with the Euler integrator ({\tt integrator = rk1}) , 
the piecewise linear method \citep{vanLeer1974JCoPh,Mignone2014JCoPh}({\tt xorder = 2}) and the Harten-Lax-van Leer Contact Riemann solvers \citep{Toro1992RSPTA,Batten1997SJSC,Toro2019ShWav} ({\tt {-}{-}flux hllc}). 
The computational range is $0 \,\mathrm{cm} \leq r \leq 4.85\times 10^{16} \, \mathrm{cm}$ discretised on a logarithmic mesh. 
We set the mesh size to $N_r=1000$ and the minimum cell width to $\Delta r_\mathrm{min} = 1.4 \times 10^7 \,\mathrm{cm}$. The computation is terminated at $10^6\,\mathrm{s}$ .
We confirm that using a second-order time integrator ({\tt integrator = vl2}) for model u18 (see Section \ref{subchap:envelope_method}) gives an almost identical fallback evolution.

We inject thermal energy to artificially drive an explosion that represents the supernova \citep[\textit{thermal bomb method;}][]{Forster2018NatAs,Sawada2019ApJ,Martinez2022A&A}. Specifically, the thermal energy $1.0 \times 10^{50} \, \mathrm{erg} \leq E_\mathrm{inj} \leq 1.0 \times 10^{52} \,\mathrm{erg}$ is injected uniformly within the innermost $10$ cells of the computational domain region ($0\, \mathrm{cm} \leq r \leq 1.1 \times 10^8 \, \mathrm{cm}$) at $t=0\,{\rm s}$. 

\subsection{Envelope Density Distribution}
\label{subchap:envelope_method}

\begin{figure}
    \centering
    \includegraphics[width=1.0\linewidth]{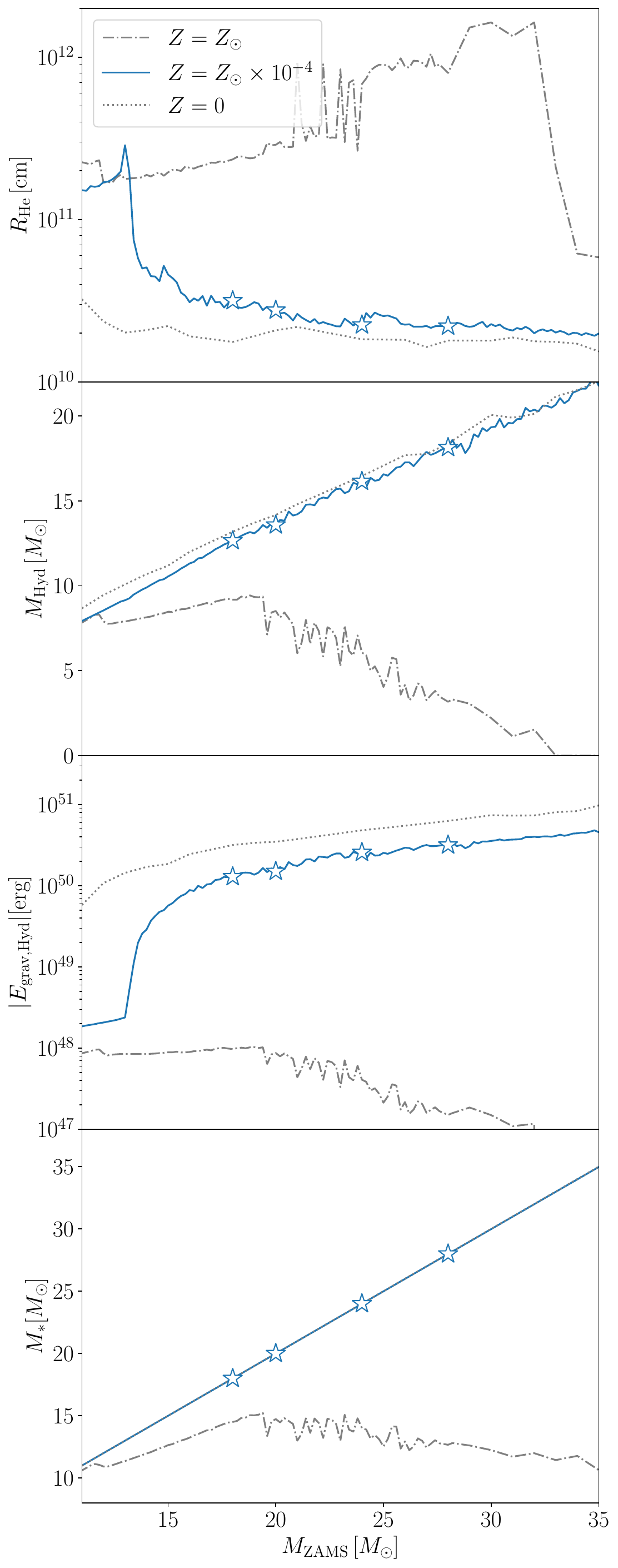}
    \caption{
    Comparison among the z- (grey dot-dash line), u- (blue line), and s-model (grey dot line). (First) For the helium core radius. (Second) For the mass of the hydrogen envelope. (Third) For the binding energy of the hydrogen envelope. (Fourth) For the final progenitor mass. The horizontal axis of each panel is $M_{\mathrm{ZAMS}}$ normalised by solar mass unit. The star symbols correspond to the progenitor we use in this study. The z-, u-, and s-model details are in the text.}
    
    \label{fig:M-Egh} 
\end{figure}

This calculation aims to perform a contrast experiment between SNII and SESN.
 This section describes a method for creating an SESN progenitor based on the stellar profile of \cite{Woosley2002RvMP}, which has the same structure as the SNII progenitor up to its helium core. 
The procedure involves first identifying the outer boundary of the helium core and then attaching a thin envelope to its surface. By doing so, we effectively remove the mass and binding energy associated with the original hydrogen envelope from the SNII progenitors.
For clarity, we label progenitor models by metallicity and initial mass as follows: solar metallicity ($Z=Z_\odot$) as s, $Z=10^{-4}Z_\odot$ as u, and $Z=0$ as z. The zero-age main-sequence mass is defined as $M_\mathrm{ZAMS}$. For example, a solar-metallicity progenitor with $M_\mathrm{ZAMS}=15\, M_\odot$ is labelled s15.

We identify the helium core surface as the radius where the gradient of the quantity $\rho r^3$ switches its sign in the boundary region between the helium core and the hydrogen envelope.
Because $\rho r^3$ is inversely correlated with the shock velocity \citep{Woosley1995ApJS}, this gradient-change point approximately corresponds to the base of the layer where the shock starts interacting with the hydrogen envelope, and it also marks the location where a reverse shock is expected to form.
We therefore adopt this radius as the effective helium-core surface to minimize the dynamical influence of the overlying envelope.

By defining the radius of the helium core, the mass and binding energy of the hydrogen envelope in the s, u, and z models can be evaluated.
The results are shown in Figure \ref{fig:M-Egh}. 
In this figure, the top panel of this figure plots the position of the helium core surface $R_\mathrm{He}$, the middle panel plots the mass of the hydrogen envelope $M_\mathrm{Hyd}$, and the bottom panel plots the absolute value of the binding energy of the hydrogen envelope $E_\mathrm{grav,Hyd}$ at each $M_\mathrm{ZAMS}$.

The three features can be identified from Figure \ref{fig:M-Egh}.
First, compared to the s model, the u models tend to have a more compact helium core and a more abundant hydrogen envelope. 
Therefore, in absolute value, the binding energy of the envelope in the u-model is also larger than that for the s-model.
Second, for the s models, $M_\mathrm{Hyd}$ decreases with increasing mass ($M_\mathrm{ZAMS} > 20 \, \mathrm{M_\odot}$), and for $M_\mathrm{ZAMS} > 32.5 \, \mathrm{M_\odot}$ the hydrogen envelope is completely lost. 
The decrease in $M_{\rm Hyd}$ also alters the relationship between $M_{\rm ZAMS}$ and the progenitor mass just before the explosion.
While $M_{*}$ is smaller than $M_{\rm ZAMS}$ because of mass loss in the s model, the two values are almost identical in the u and z models.
Third, the $M_\mathrm{Hyd}$ of the u model monotonically increases, while $R_\mathrm{He}$ decreases. 
Hence, 
$|E_\mathrm{grav,Hyd}|$ monotonically increases with $M_{\rm ZAMS}$.
We choose the progenitors of the u models with large binding energy of the hydrogen envelope for the initial conditions in order to elucidate the impact of the hydrogen envelope.

Naturally, if we simply cut off the density and pressure at $R_\mathrm{He}$, the structure of the progenitor's envelope is out of hydrostatic equilibrium. In this calculation, we use the method of \cite{Matzner1999ApJ} to attach a thin outer shell that maintains hydrostatic equilibrium from the surface of the helium core. The radial distribution of density is given as follows,
 \begin{align}
    \label{layer_connect}
    \rho = \begin{cases}  \rho_0 \displaystyle\left(\frac{R_*}{r} - 1\right)^{n_{\mathrm{He}}} &\quad \rho_1< \rho\leq\rho_0 \\ \rho_1\displaystyle\left(\frac{r_1}{r}\right    )^2 &\quad \rho \leq \rho_1 \end{cases}.
\end{align}

The upper branch of Equation \ref{layer_connect} represents the isentropic helium core, and the lower branch corresponds to the stellar wind region. 
The isentropic assumption applies only between the core surface ($\rho_0$) and the transition density $\rho_1 = 1.0 \times 10^{-7}\,\mathrm{g\,cm^{-3}}$.
The index $n_\mathrm{He}$ is obtained by matching the density profile to two adjacent grid points of the progenitor at the core surface, providing the condition to close Equation \ref{layer_connect}.
The pressure there follows a polytropic relation $p = K\rho^\gamma$, where $K = p_0’ / \rho_0’^{\gamma}$ is derived from the innermost cell just before truncation.
This profile approximately satisfies hydrostatic equilibrium and any deviation is negligible over the calculation time. 
For $\rho \le \rho_1$, the density follows the prescribed wind profile, which is not in hydrostatic equilibrium but smoothly connects to the core structure.

\begin{figure}
    \centering
    \includegraphics[width=1.\linewidth]{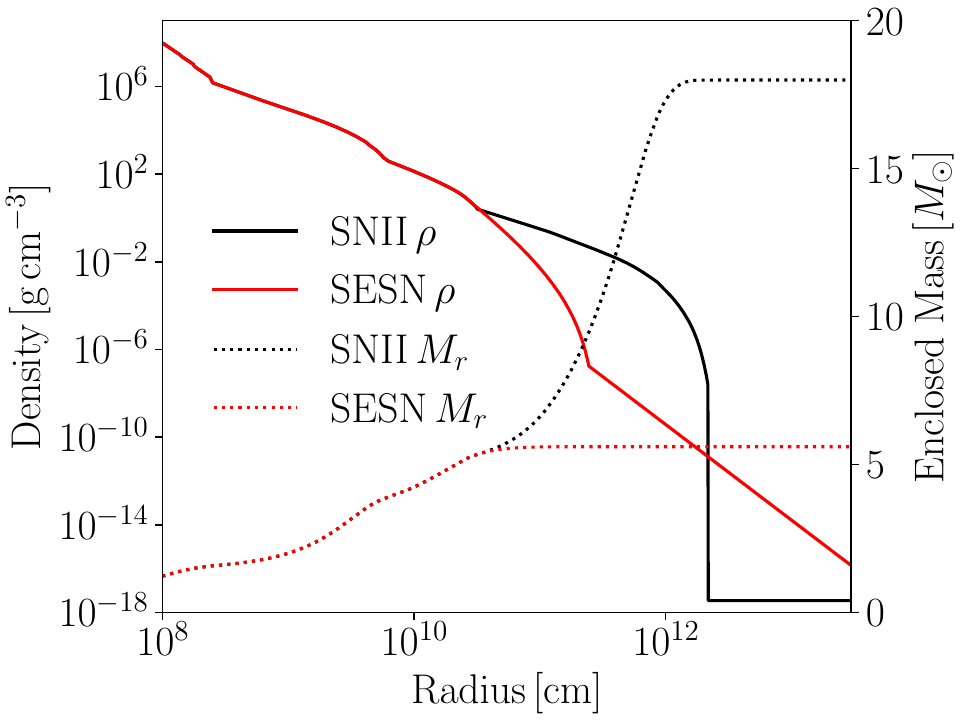}
    \caption{The density profile (solid line) and enclosed mass profile (dotted line) of $M_\mathrm{ZAMS}=18\, \mathrm{M_\odot},\,Z=10^{-4}\, \mathrm{Z_\odot}$ (u18 model) from centre to surface. The hydrogen rich progenitor (SNII)  are shown in black and hydrogen poor progenitor (SESN) in red. Left y-axis shows density and right y-axis shows enclosed mass in solar mass unit.}
    \label{fig:u18_r_rho_Men}
\end{figure}

The radial distribution of density (solid line) and enclosed mass (dotted line) are plotted in Figure \ref{fig:u18_r_rho_Men}. 
The SNII progenitor in this figure is the stellar profile from \cite{Woosley2002RvMP}.
The density and enclosed mass structures are consistent up to the helium core surface position $R_\mathrm{He} \simeq 3.2 \times 10^{10} \mathrm{cm}$ in the u18 model, and the enclosed mass of the SESN progenitor is constant after that radius.

\subsection{Core Density Distribution : core softening method}
\label{subchap:softened_method}

Due to the steep density gradients and the high sound speed near the neutron star,  there are difficulties in treating the innermost stellar regions.
Conventional studies have adopted sink boundaries that exclude the neutron star from the computational domain \citep{MacFadyen2001ApJ, Zhang2008ApJ, LoveGrove2013ApJ, Dexter2013ApJ, Fernandez2018MNRAS}.
However, this approach may not accurately capture the incoming reverse shock, producing artificial waves and resulting in underestimated fallback accretion rates and remnant masses \citep{Ertl2016ApJ, Gabler2021MNRAS, Sykes2025MNRAS}.

To avoid this artefact, we adopt the {\it core softening method} \citep{Hirai2020MNRAS}.
In this method, the stellar core is replaced with a point-like particle surrounded by a thin, low-density sphere, 
thereby modifying the density and gravitational potential in the innermost region without relying on the uncertain choices of conventional inner boundary conditions.
The potential of the point-like particle is softened to avoid singularities \citep[e.g.,][]{Ohlmann2017A&A}, and the enclosed point mass $M_\mathrm{pt}$ is uniquely determined by the softening radius $r_s$.

A cubic spline is used for the softened gravitational potential of the point source particle, ${\phi_\mathrm{pt}}$ \citep[equation A2 in][]{Price2007MNRAS}.
An artificial hydrostatic distribution is constructed for the gas in the softened region $(r < r_s)$. The four equations that the softened density and pressure distribution $(\rho_s, p_s)$ must satisfy are as follows,
\begin{align}
    & \nabla p_\mathrm{s}(r)+G\rho_\mathrm{s}(r)\left(\frac{m_\mathrm{s}(r)}{r^2}+M_\mathrm{pt}\nabla\phi_\mathrm{pt}(r,r_\mathrm{s})\right)=0,\label{eq:hydrostatic}\\
    & p_\mathrm{s}(r_\mathrm{s})=p(r_\mathrm{s}),\\
    &S_{\rm s} (r) = S(r_{\rm s}), \\
    & \nabla p_\mathrm{s}(r_\mathrm{s})=\nabla p(r_\mathrm{s}),\\
    & M_\mathrm{pt}+m_\mathrm{s}(r_\mathrm{s})=m(r_\mathrm{s}).\label{eq:mass_constraint}
\end{align}

Here, $\rho (r)$, $m(r)$, and $S(r)$ are the density, mass coordinates, and entropy of the original stellar profile, and $\rho_\mathrm{s} (r)$, $m_\mathrm{s}(r)$, and $S_{\rm s}(r)$ are quantities of the softened profile. $m_s (r)$ is defined as follows,
\begin{align}
    \label{eq:def_mass_soft}
    m_\mathrm{s}(r)\equiv4\pi\int_0^r\rho_\mathrm{s}(r')r'^2dr'.
\end{align}

\begin{figure}
\centering
\includegraphics[width=1.\linewidth]{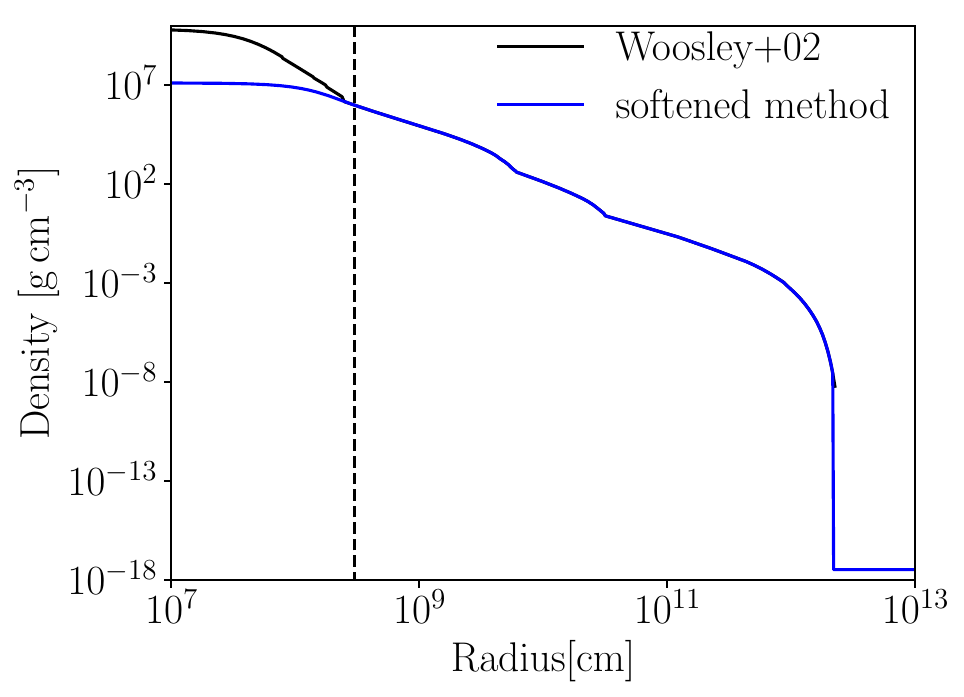}
\caption{The density profile before (black solid line) and after (blue solid line) softening of $M_\mathrm{ZAMS}=18\, \mathrm{M_\odot},\,Z=10^4\, \mathrm{Z_\odot}$ (u18 model). We set softening radius to $3.0 \times 10^8\, \mathrm{cm}$ (black dashed line).}
\label{fig:WHW02_vs_Hirai_density}
\end{figure}

For the u18 model, we choose a value of $r_s = 3.0 \times 10^8 , \mathrm{cm}$, corresponding to a mass coordinate $M_{r_s} \simeq 1.6,{\rm M_\odot}$ (see Figure \ref{fig:WHW02_vs_Hirai_density}).
This radius is chosen so that the point mass roughly corresponds to $M_{\mathrm{pt}} \simeq 1.5,{\rm M_\odot}$ in the u18 model, and the same radius ($r_s = 3.0 \times 10^8\,\mathrm{cm}$) is adopted for the other progenitor models.
After solving for the above equations, the mass of the point source particle is about $1.48 \, \mathrm{M_\odot}$ and is about the mass of a neutron star.
The progenitor used in the explosion calculations, $M_\mathrm{pt}$, are noted in Table \ref{tab:prog_val}.
In this work, we choose $r_s = 3 \times 10^8 \, {\rm cm}$ for all progenitors. To check the dependence on this value, we also perform explosion calculations with $r_s = 1 \times 10^9 \, {\rm cm}$ in u18, and we find that there is no significant difference.

$M_{\rm pt}$ is not changed during the explosion calculation. In other words, $M_{\rm pt}$ is the {\it computationally} minimum mass of the central compact object that forms immediately after the explosion. In this calculation, our goal is for a neutron star, rather than a black hole, to form initially.

\begin{table*}
    \centering
    \caption{
    The table shows 
    point mass$(M_\mathrm{pt})$, mass of hydrogen envelope $(M_\mathrm{Hyd})$, the radius of He core surface $(R_\mathrm{He})$ , the binding energy of hydrogen envelope $(E_\mathrm{grav,Hyd})$,
    mass of C/O core $(M_\mathrm{CO})$, the radius of C/O core surface $(R_\mathrm{CO})$, and the ratio of 
    of each progenitors.}
    \begin{tabular}{cccccccc}
        \hline
         progenitor 
         &$M_\mathrm{pt} \,[\, \mathrm{M_\odot}]$ &   $M_\mathrm{Hyd}\, [\, \mathrm{M_\odot}]$ & $R_\mathrm{He}\, \mathrm{[cm]}$ & $E_\mathrm{grav,Hyd} \,\mathrm{[erg]}$ &
         $M_{\rm CO} \, [\rm M_\odot]$ &
         $R_{\rm CO} \, [\rm cm]$ & $\dfrac{M_{\rm CO}/{\rm M_\odot}}{R_{\rm CO}/10^9{\rm cm}}$\\
\hline
  u18 & 
  $1.48$  &  $12.7$  &  $3.17\times 10^{10}$ &  $-1.29\times 10^{50}$ 
  &$3.69$ & $4.80\times10^9$ & $0.77$
  \\
   u20 
   &  $1.63$  &  $13.6$  &  $2.77\times 10^{10}$ &  $-1.50\times 10^{50}$ 
  &$4.65$ & $3.89\times10^9$ & $1.19$
   \\
   u24 & 
   $1.26$  &  $16.2$  &  $2.24\times 10^{10}$ &  $-2.57\times 10^{50}$ 
  &$6.10$ & $4.64\times10^9$ & 
  $1.31$
   \\
   u28 &  
   $1.65$  &  $18.1$  &  $2.21\times 10^{10}$ &  $-3.17\times 10^{50}$ 
  &$7.84$ & $5.79\times10^9$ & $1.35$
   \\
\hline
    \end{tabular}
    \label{tab:prog_val}

\end{table*}

Note that,  a cooling term,
\begin{align}
    Q_\mathrm{cool} = 
    \begin{cases}
        \frac{\varepsilon(r) - \varepsilon_*(r)}{\tau_\mathrm{cool}}  & \quad r \leq r_\mathrm{s}\\
        0 & \quad r > r_\mathrm{s}
    \end{cases},
\end{align}
is introduced to avoid the effect of high entropy of the initial thermal bomb on fallback after the explosion, where we set the cooling timescale, $\tau_\mathrm{cool} = 30 \,\mathrm{s}$. Here, $\varepsilon(r)$ denotes the internal energy density at position $r$. The $\varepsilon_*(r)$ is the internal energy density reflecting the entropy structure of the progenitor before the injection of the thermal bomb. Specifically, it is written as,
\begin{align}
    \varepsilon_*(r) = \varepsilon_\mathrm{init} 
    \qty(\frac{\rho (r)}{\rho_\mathrm{init} (r)})^\gamma,
\end{align}
where $\rho_\mathrm{init}(r),\,\varepsilon_\mathrm{init}(r)$ are respectively the density and internal energy at position $r$ before the thermal bomb is injected.

\section{RESULTS}
\label{sec:results}
\subsection{Shock Propagation, Fallback Accretion Rate and Remnant Mass }
\begin{figure}
    \centering
    \includegraphics[width=1.0\linewidth]{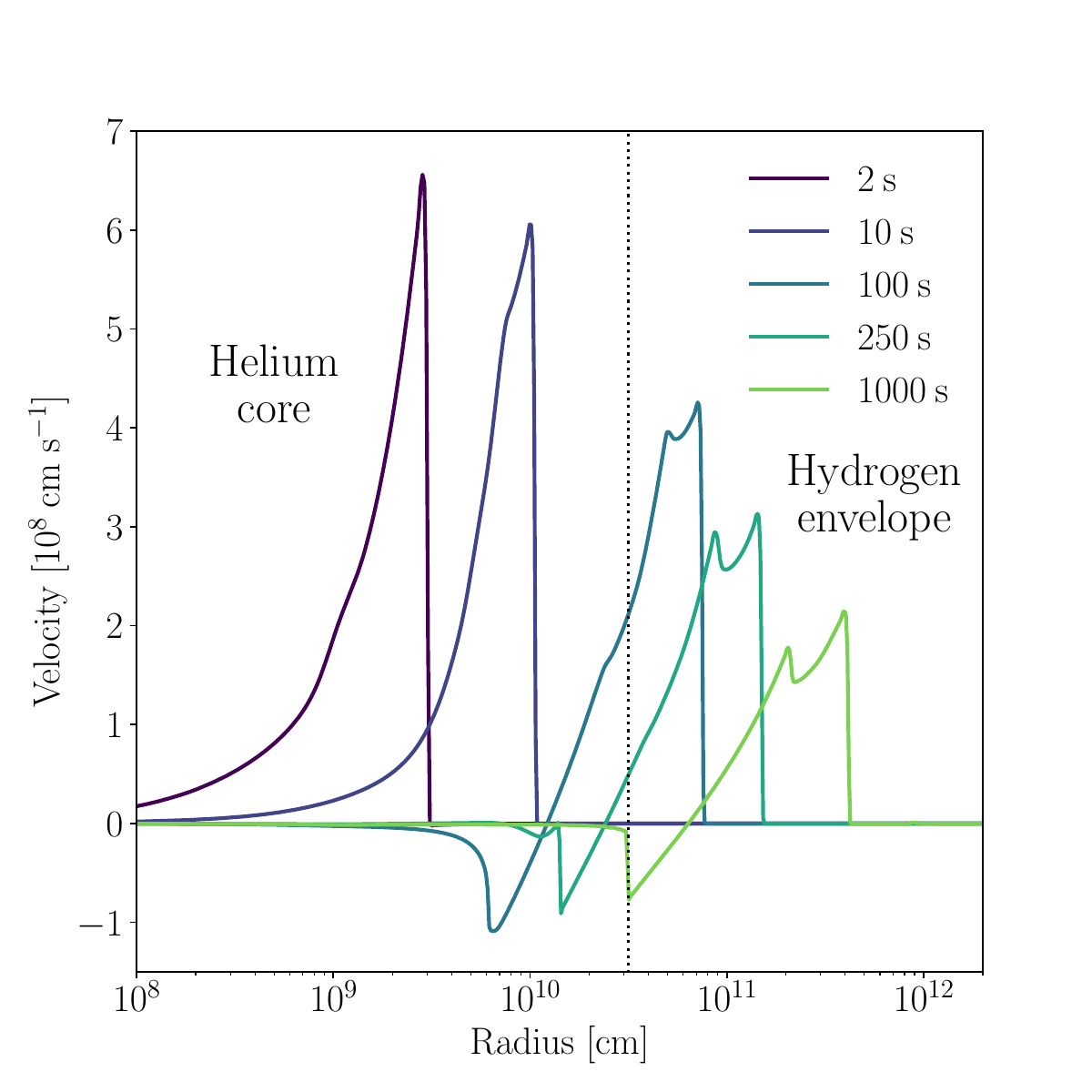}
    \caption{Snapshots of the time evolution of the shock wave's velocity propagating through the progenitor's interior with a hydrogen envelope (SNII). The time for each curve is shown in the top right of the figure. The horizontal axis is the radius, and the vertical axis is the velocity. The dotted line represents the radius of the helium core surface.}
    \label{fig:u18_vel_snap}
\end{figure}

Figure \ref{fig:u18_vel_snap} shows the time evolution of the velocity of a shock wave propagating through the interior of the progenitor. The shock wave propagating through the helium core (left of the dotted line) has only one peak ($t=2$, $10\,\mathrm{s}$), but when it propagates through the hydrogen envelope (right of the dotted line), the shape of this shock wave changes and has two peaks ($t = 100 $, $250$, $1000\,\mathrm{s}$). From here on, we call them the {\it forward shock} and the {\it reverse shock}.

The shock wave forms a double-peaked structure as it encounters the large $\rho r^3$ of the hydrogen envelope.
An increasing $\rho r^3$ rapidly decelerates the forward shock velocity \citep{Woosley1995ApJS} and this generates a reverse shock which moves inward from the forward shock coordinate system.

On the other hand, the shock wave with a negative velocity behind the reverse shock (hereafter, an {\it accretion shock}) occurs when $t>100\,\mathrm{s}$. This shock is formed by the collision between the accretion flow and the equilibrium atmosphere on the central compact object that has accumulated due to accretion.

	\begin{figure*}
    \centering
    \includegraphics[width=1.\linewidth]{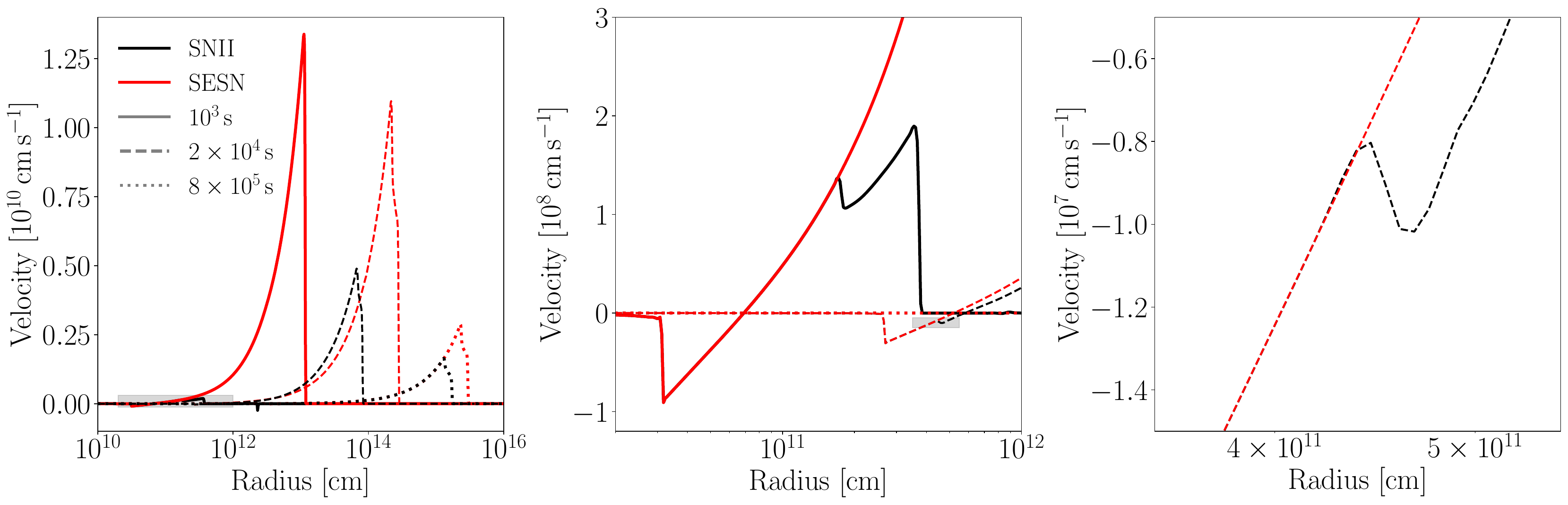}
    \caption{
    Time evolution of the radial distribution of the shock velocity for progenitors with (SNII, black) and without hydrogen envelope (SESN, red). (Left): Overview of the shock wave structure. (Middle): Snapshot focusing on the early formation of the reverse shock. (Right): Snapshot highlighting the reverse shock during falling. We inject the same thermal bomb energy, $E_\mathrm{inj} = 8.0\times 10^{50} , \mathrm{erg}$, into both progenitors. The solid, dashed, and dotted lines correspond to the distributions at $10^3\,\mathrm{s}$, $ 2\times10^4\,\mathrm{s},$ and $8\times 10^5\,\mathrm{s}$ respectively, after the energy injection. Note that the gray areas in the left and middle panels indicate the drawing areas of the middle and right panels, respectively.}
    \label{fig:u18_8e50_vel}
\end{figure*}

Figure \ref{fig:u18_8e50_vel} shows the temporal evolution of the shock waves in the progenitors of different types.
This figure compares the radial distribution of the shock velocity in the progenitor with a hydrogen envelope (SNII, black) and that without (SESN, red) to check how the shock wave structure evolves. The solid, dashed, and dotted lines correspond to times of $10^3\,\mathrm{s}, 2\times10^4\,\mathrm{s}$ and $8\times 10^5\,\mathrm{s}$, respectively. The details of each panel are discussed below.

The left panel shows the global evolution of the shock waves.
After the injection of thermal energy, the shock wave propagates outward and decelerates over time. In the SNII progenitor, the hydrogen-rich envelope  significantly reduces the maximum velocity of the forward shock. In contrast, in the SESN progenitor, the density outside the helium core ($r > R_\mathrm{He}$, see Figure \ref{fig:u18_r_rho_Men}) is lower than in the SNII progenitor, resulting in a sharp increase in the shock velocity. This phenomenon is attributed to the inverse correlation between $\rho r^3$ and the maximum shock velocity, as discussed in \cite{Woosley1995ApJS}.

The middle panel shows the initial formation of the reverse shock.
At $10^3$ and $2\times10^4\,\mathrm{s}$, the shock wave in the SNII progenitor interacts with the hydrogen envelope, forming a reverse shock. In contrast, the SESN progenitor lacks a hydrogen envelope and steep $\rho r^3$ gradient. As a result, no reverse shock forms, and the rapid increase in the fallback accretion seen by \cite{Zhang2008ApJ} does not occur in our SESN case.

\begin{figure*}
    \centering
    \includegraphics[width=1.0\linewidth]{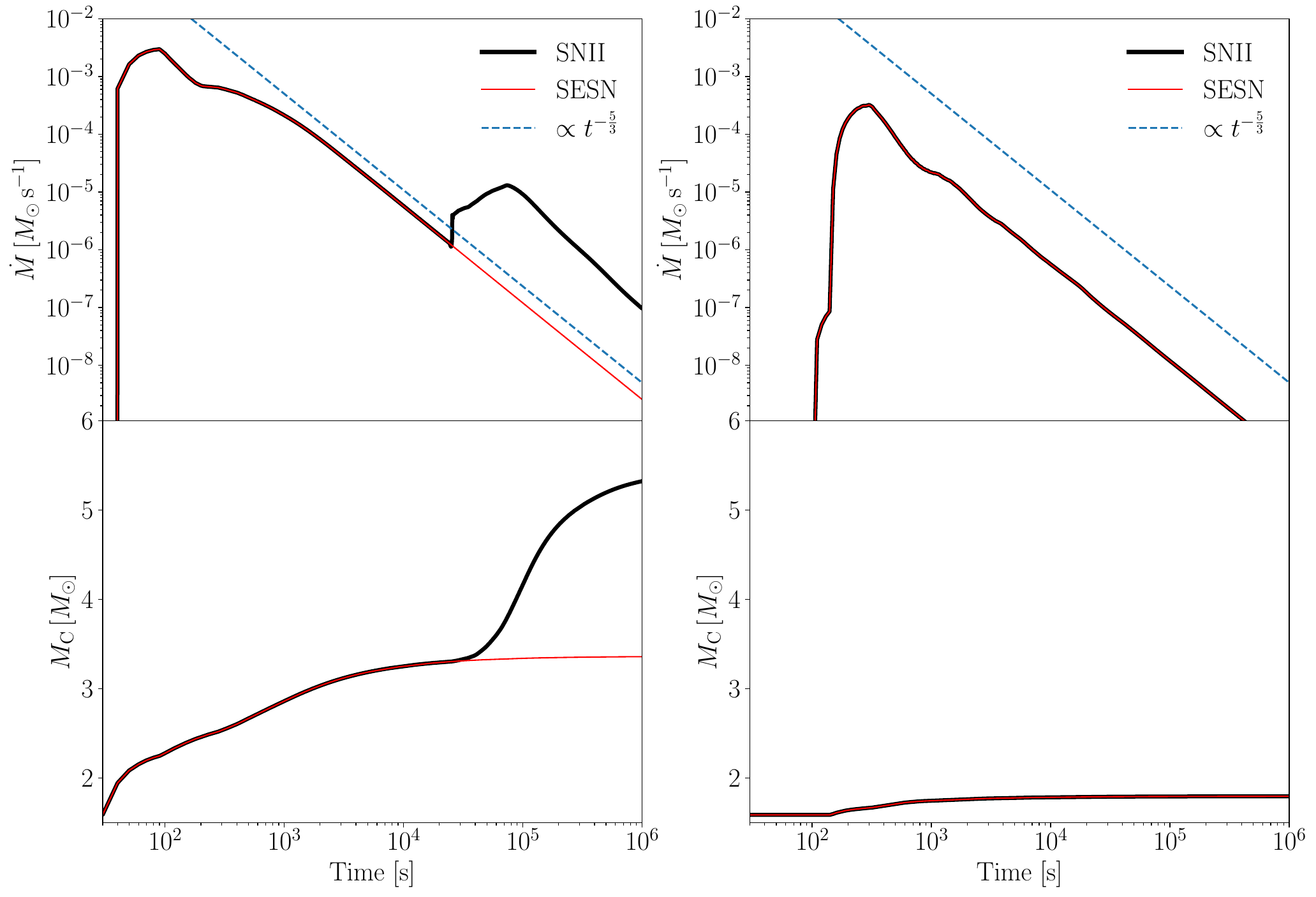}
    \caption{The upper and lower panels show the time evolution of the accretion rate and the remnant mass for the metallicity $10^{-4}\, \mathrm{Z_\odot}$, $M_\mathrm{ZAMS} = 18 \, \mathrm{M_\odot}$ and the thermal bomb energy ($E_\mathrm{inj}$) of $8\times10^{50}  \,\mathrm{erg}$ (left) and $2\times 10^{51}\,\mathrm{erg}$ (right). The black line shows the case of the progenitor with a hydrogen envelope (SNII), and the red line shows the case of the progenitor without a hydrogen envelope (SESN). The blue dashed line shows the asymptotic accretion rate ($\propto t^{-5/3}$). In the case of $E_\mathrm{inj} = 2 \times 10^{51}\,\mathrm{erg}$, the reverse shock is directed outward until the end of the calculation time of $t=10^6\, {\rm s}$ and its velocity exceeds the escape velocity.
    }
    \label{fig:u18_t_Mdot_Mc}
\end{figure*}

The right panel shows the reverse shock during fallback accretion.
At $8\times 10^5\,\mathrm{s}$, the reverse shock in the SNII progenitors propagates back toward the central compact object, triggering a sharp increase in fallback accretion and influencing the remnant mass evolution. In the SESN progenitors, however, the absence of reverse shock results in continued accretion following the free-expansion solution (see Figure \ref{fig:u18_t_Mdot_Mc}).

Figure \ref{fig:u18_t_Mdot_Mc} shows the time evolution of the accretion rate and the remnant mass of u18 whose $E_\mathrm{inj}$ are $ 8.0\times 10^{50}\,\mathrm{erg}$ (left) and $2.0\times10^{51}\,\mathrm{erg}$ (right), respectively. 
The accretion rate is defined as
\begin{align} 
\dot{M} = -4\pi  v_{\rm min} r_{v_{\rm min}}^2 \rho(r_{v_{\rm min}}),
\end{align}
where $v_{\rm min}$ is the radial velocity and $r_{v_{\rm min}}$ is the radius at which the radial velocity takes its minimum value. $r_{v_{\rm min}}$ corresponds to the location of the accretion shock. The remnant mass is defined as $M_{\rm C} = M_r(r_{v_{\rm min}}).$

Focusing on the left panel of SNII (black line), where the $E_{\rm inj}$ is small and the reverse shock falls, three main stages can be distinguished.
First, the accretion rates in both models decrease asymptotically, following a $t^{-\frac{5}{3}}$ dependence, as in the free expansion \citep{Chevalier1989ApJ}.
Second, the reverse shock of SNII progenitor (black line) falls to the centre at $2.5\times 10^4 \,\mathrm{s}$, resulting in a significant increase in accretion and remnant mass. 
Thirdly, the decline rate of the accretion returns to the analytical solution of the free expansion phase.
At the same $E_{\rm inj}$ as SNII (the left panel, red line), however, the accretion rate of SESN decreases following $t^{-\frac{5}{3}}$ for the entire duration because there is no Hydrogen envelope of this progenitor.

In the right panel, however, $E_{\rm inj}$ is sufficiently large, and the reverse shock of SNII escapes the gravitational potential of the remnant mass. Therefore, the time evolution of the accretion rate and remnant mass of both SNII and SESN are identical.

\begin{figure}
    \centering
    \includegraphics[width=1.0\linewidth]{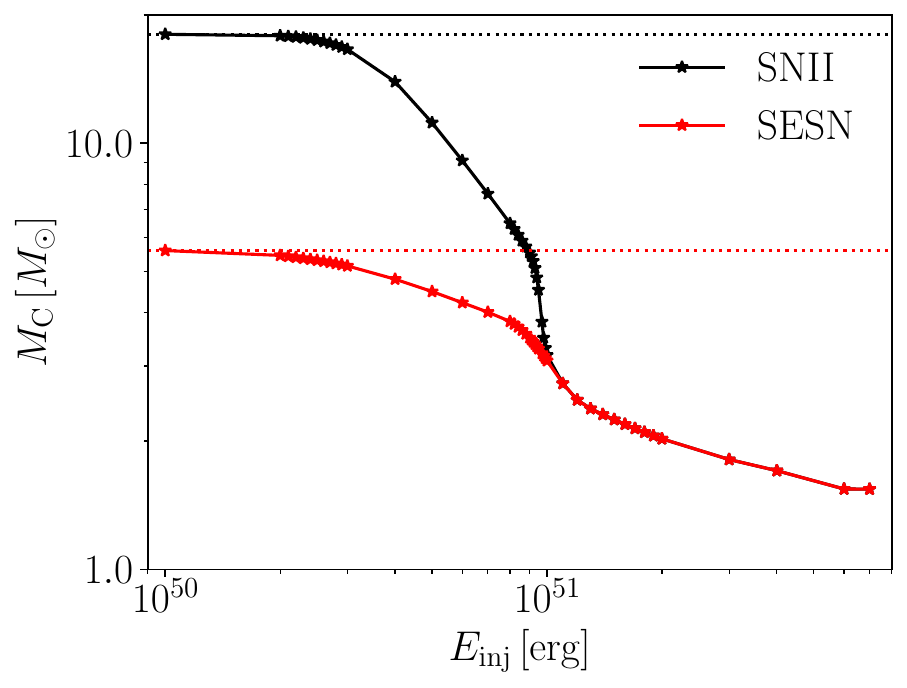}
    \caption{Thermal bomb energy $(E_\mathrm{inj})$ and remnant mass $(M_\mathrm{C})$ for the progenitor with hydrogen envelope (SNII, black line) and the progenitor without hydrogen envelope (SESN, red line) at metallicity $10^{-4}\, \mathrm{Z_\odot}$ and $M_\mathrm{ZAMS} = 18 \, \mathrm{M_\odot}$. The dotted lines are the total mass of each progenitors. Each point corresponds to the calculated model.}
    \label{fig:u18_Etb_MC}
\end{figure}

Figure \ref{fig:u18_Etb_MC} shows the relation between thermal bomb energy and compact object mass. It can be seen that for $E_\mathrm{inj}\gtrsim1.0\times10^{51}\,\mathrm{erg}$, both SNII and SESN leave the same remnant mass, while they draw different trajectories at smaller energies. 
Interestingly, there is a steep jump in the compact remnant mass at around the region where the thermal bomb energy is $8.0\times 10^{50}\, \mathrm{erg} \lesssim E_\mathrm{inj} \lesssim 1.0\times10^{51}\, \mathrm{erg}$ in the SNII case. Hereafter, we call this 
the mass {\it transition}.
The SESN case shows a similar {\it transition} region but with a smaller mass variation than in the SNII case.
Therefore, this figure can be divided into three major regions.

(1) High energy side ($E_\mathrm{inj} \gtrsim 9.0\times 10^{50} \, \mathrm{erg}$): The trajectories of SNII and SESN cases are coincident,  because the structures of both progenitor types are coincident except for the hydrogen envelope, and the reverse shock wave generated in the hydrogen envelope of the SNII cases does not fall into the centre.

(2) Intermediate range ($8.0\times 10^{50}\, \mathrm{erg} \lesssim E_\mathrm{inj} \lesssim 9.0 \times 10^{50}\, \mathrm{erg}$): 
In this range, the effect of the hydrogen envelope appears.
In order to resolve the rapid change in $M_{\rm c}$, we sample the thermal bomb energy more finely than in the other energy ranges. 
Note that the {\it transition} region appears in the SNII trajectory due to the reverse shock wave generated in the hydrogen envelope. 
Consequently, there is a jump in remnant mass between $M_{\rm C} = 3.0\, \mathrm{M_\odot}$ and $5.7\, \mathrm{M_\odot}$ which encompasses the helium layer in the progenitor. This indicates that the helium layer can be ejected at almost no additional cost of energy as long as enough energy was injected to eject the overlying hydrogen envelope. 
In contrast, for SESNe, the jump is much smaller and therefore connects more smoothly to the high-energy side. This highlights the importance of the reverse shock, as the increase in ejected helium mass is more gradual in the SESN case, while there is an almost binary outcome depending on whether the reverse shock can fall back or not in the SNII case.

(3) Low energy side ($E_\mathrm{inj} \lesssim 8.0 \times 10^{50}\,\mathrm{erg}$): The trajectories of both types do not coincide, and SNII constantly forms a heavier compact object than SESN. Because the reverse shock wave formed in the hydrogen envelope falls into the centre, in contrast to the case (1), the response of the compact object mass to the energy of SNII is steeper than that of SESN.

On the observational side, we cannot directly get the thermal bomb's energy. Thus, in this section, we discuss the remnant mass distribution using the observed quantity of the explosion energy,  $E_{\rm exp}$, which is obtained by calculating the total energy of the ejecta at the end of the simulation,
\begin{align}
    E_\mathrm{exp} = \int_{v>v_\mathrm{esc}}e_\mathrm{tot}dV,
\end{align}
where $v_\mathrm{esc}$ is escape velocity, and $e_\mathrm{tot}$ is the sum of internal, kinetic, and gravitational energy density.

\begin{figure}
    \centering
    \includegraphics[width=1.\linewidth]{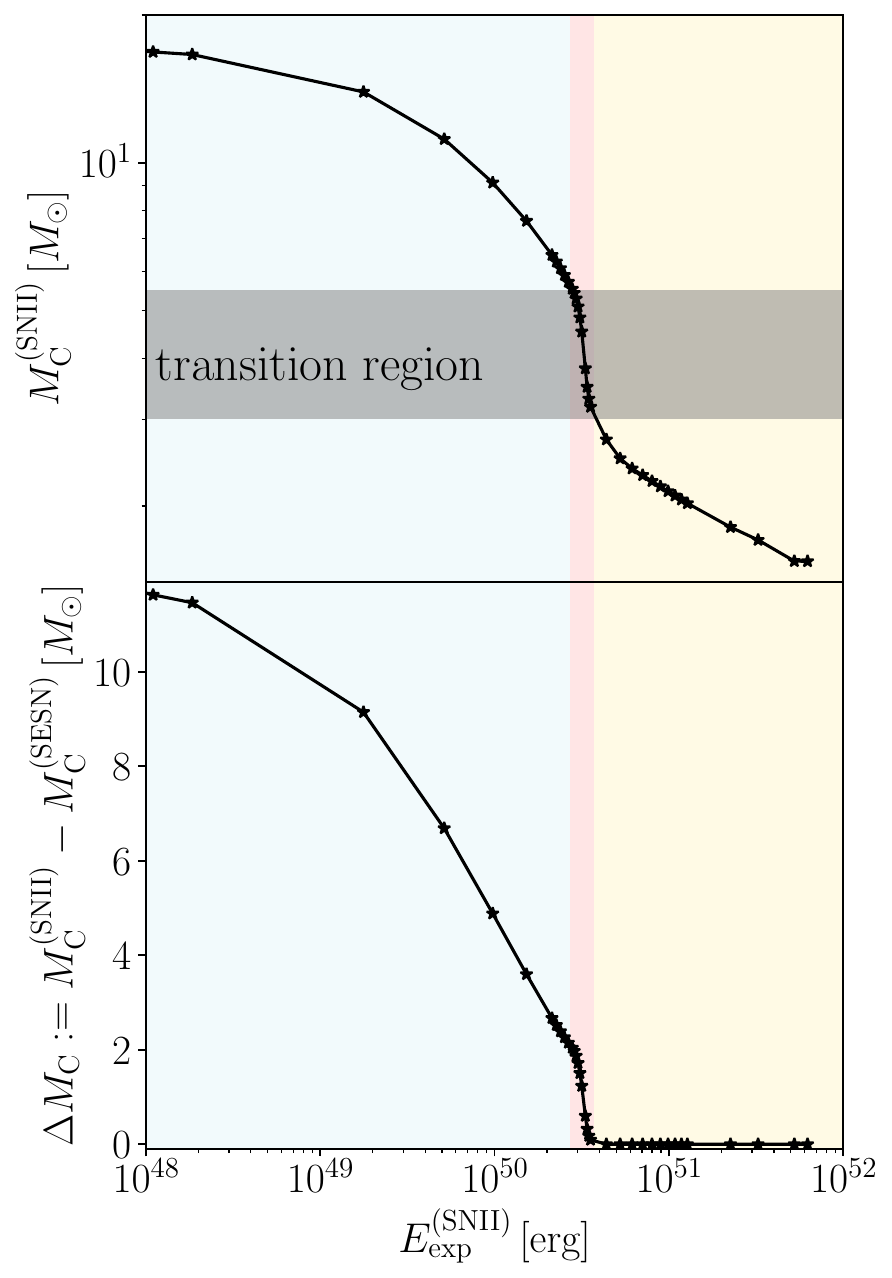}
    \caption{The explosion energy and remnant mass of the progenitor with hydrogen envelope (top panel), and the difference in the remnant masses for models with and without hydrogen envelope, $\Delta M_\mathrm{C}$ (lower panel). The horizontal axis of both panels is the explosion energy of the progenitor with hydrogen envelope. Both used the same progenitor at metallicity $10^{-4}\, \mathrm{Z_\odot}$ and $M_\mathrm{ZAMS} = 18 \, \mathrm{M_\odot}$.
    Note that the horizontal axis is not the thermal bomb energy but the explosion energy. 
    The red area corresponds to the range of explosion energy $8\times10^{50} \, \mathrm{erg} \leq E_\mathrm{inj} \leq 9\times10^{50}\,\mathrm{erg}$, the blue area corresponds to the lower energy, and the yellow area corresponds to the higher explosion energy. The range of remnant mass corresponding to the red explosion energy range is shown in the top panel as the {\it transition} region.}
    \label{fig:u18_Eex_MC_dM}
\end{figure}

The upper panels of Figure \ref{fig:u18_Eex_MC_dM} and Figure \ref{fig:u18_Etb_MC}  is almost same but the horizontal axis of Figure \ref{fig:u18_Eex_MC_dM} has changed from $E_{\rm inj}$ to $E_{\rm exp}$. And the intermediate region in Figure \ref{fig:u18_Etb_MC} has shifted to $2.5 \times 10^{50} \mathrm{erg} \lesssim E_\mathrm{exp} \lesssim 4 \times 10^{50} \mathrm{erg}$.
The lower panel shows the relationship between the explosion energy of SNII and the difference in remnant mass between SNII and SESN, $\Delta M_C := M_\mathrm{C}^\mathrm{(SNII)} - M_\mathrm{C}^\mathrm{(SESN)}$ in u18.
This value exhibits behaviour qualitatively similar to $M_C$, tending towards the hydrogen envelope mass at the low-energy limit and approaching zero at the high-energy region. At the intermediate region,  the value of $\Delta M_\mathrm{C}$ in the lower panel decreases from zero to about $2 \, \mathrm{M_\odot}$.

\subsection{Parameter Search and Normalization}

\begin{figure}
    \centering
    \includegraphics[width=1.05\linewidth]{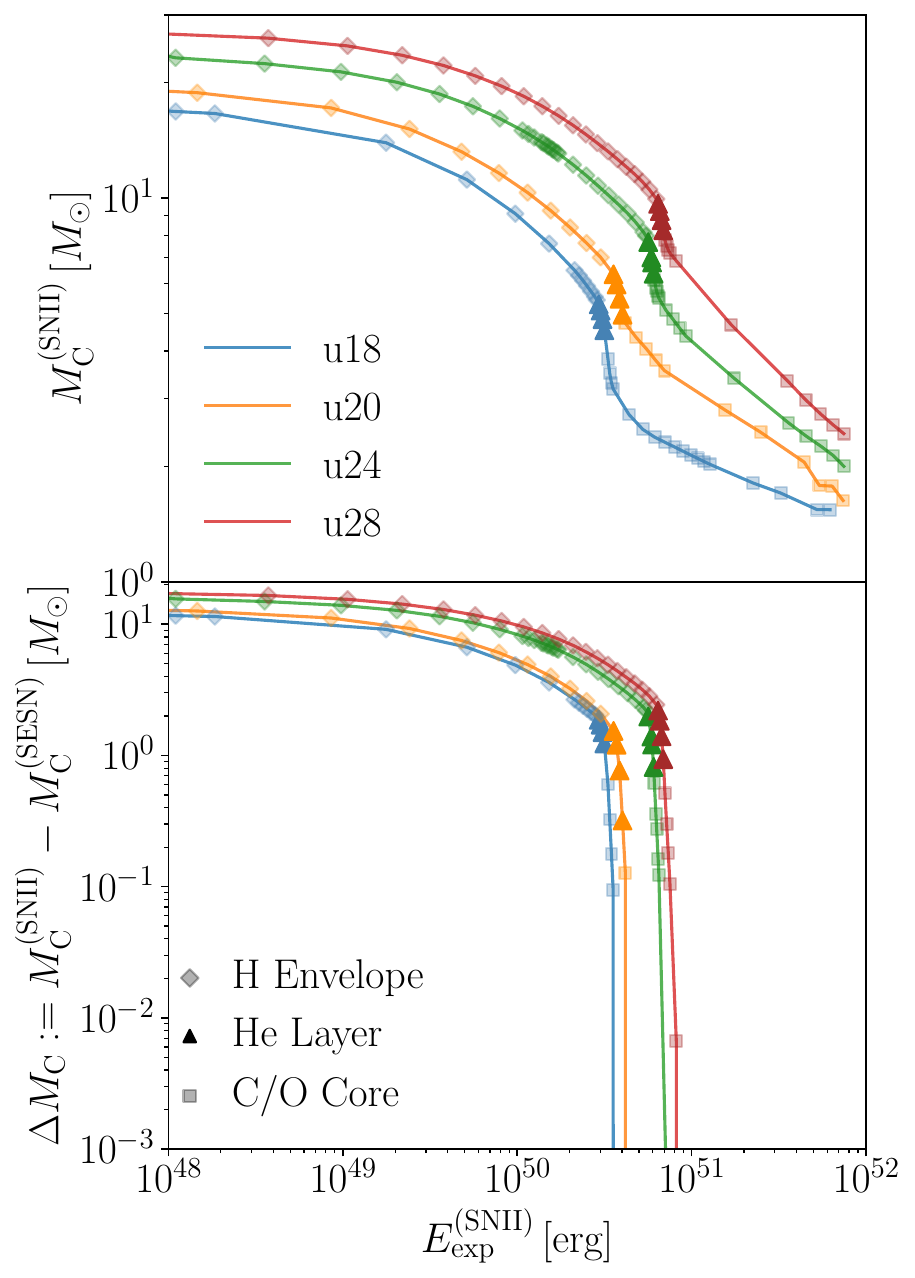}
    \caption{Similar to Figure \ref{fig:u18_Eex_MC_dM}, but for $M_\mathrm{ZAMS}=18\, \mathrm{M_\odot}$(blue), $20\, \mathrm{M_\odot}$ (orange), $24\, \mathrm{M_\odot}$(green), $28\, \mathrm{M_\odot}$(red) at metallicity $10^{-4}\, \mathrm{Z_\odot}$. In addition, the mass regions of each progenitor's hydrogen envelope, helium layer, and C/O core are represented by diamonds, triangles, and squares, respectively. Note that the vertical axis of the lower panel is logarithmic notation to clarify the {\it transition} region.
    }
    \label{fig:eachprog_Eex_MC_dM}
\end{figure}

Figure \ref{fig:eachprog_Eex_MC_dM} is the result of adding $M_\mathrm{ZAMS} = 20\, \mathrm{M_\odot}$, $24\, \mathrm{M_\odot}$, $28 \, \mathrm{M_\odot}$ to Figure \ref{fig:u18_Eex_MC_dM} (hereafter, we call these as u20, u24, u28). To clarify the relationship between the presupernova progenitor’s structure and the remnant mass, we add the mass coordinates of each compositional layer  
(Hydrogen envelope : diamond, He layer : triangle, C/O core : square). 
The He layer is defined as the region between the He core and the C/O core, with their boundary set at $X_\mathrm{He} > 0.01$.

One can clearly see that the trajectories of explosion energy and remnant mass shift toward larger masses and higher energies with larger $M_{\rm ZAMS}$ (top panel).
This is because the He core and hydrogen envelope of the progenitor generally become more tightly bound as the $M_\mathrm{ZAMS}$ increases (Figure \ref{fig:M-Egh}).

We find that the transition regions exist not only for u18 but also for other progenitors. Moreover, our results indicate that the upper (lower) side of the transition region corresponds to the He layer (C/O core) of the progenitor.
Note that the lower panel of Figure \ref{fig:eachprog_Eex_MC_dM} is similar to the lower panel of Figure \ref{fig:u18_Eex_MC_dM}, but we set the vertical axis in logarithmic scale. 
Specifically, for $\Delta M_\mathrm{C} \lesssim 0.5 {\rm M_\odot}$, only the C/O core falls back onto the compact object. Once $\Delta M_\mathrm{C}$ exceeds $\sim0.5 {\rm M_\odot}$, the He layer also begins to fall back. Furthermore, when $\Delta M_\mathrm{C} \gtrsim 3{\rm M_\odot}$, even the hydrogen envelope starts to be accreted.

\begin{figure}
    \centering
    \includegraphics[width=1.\linewidth]{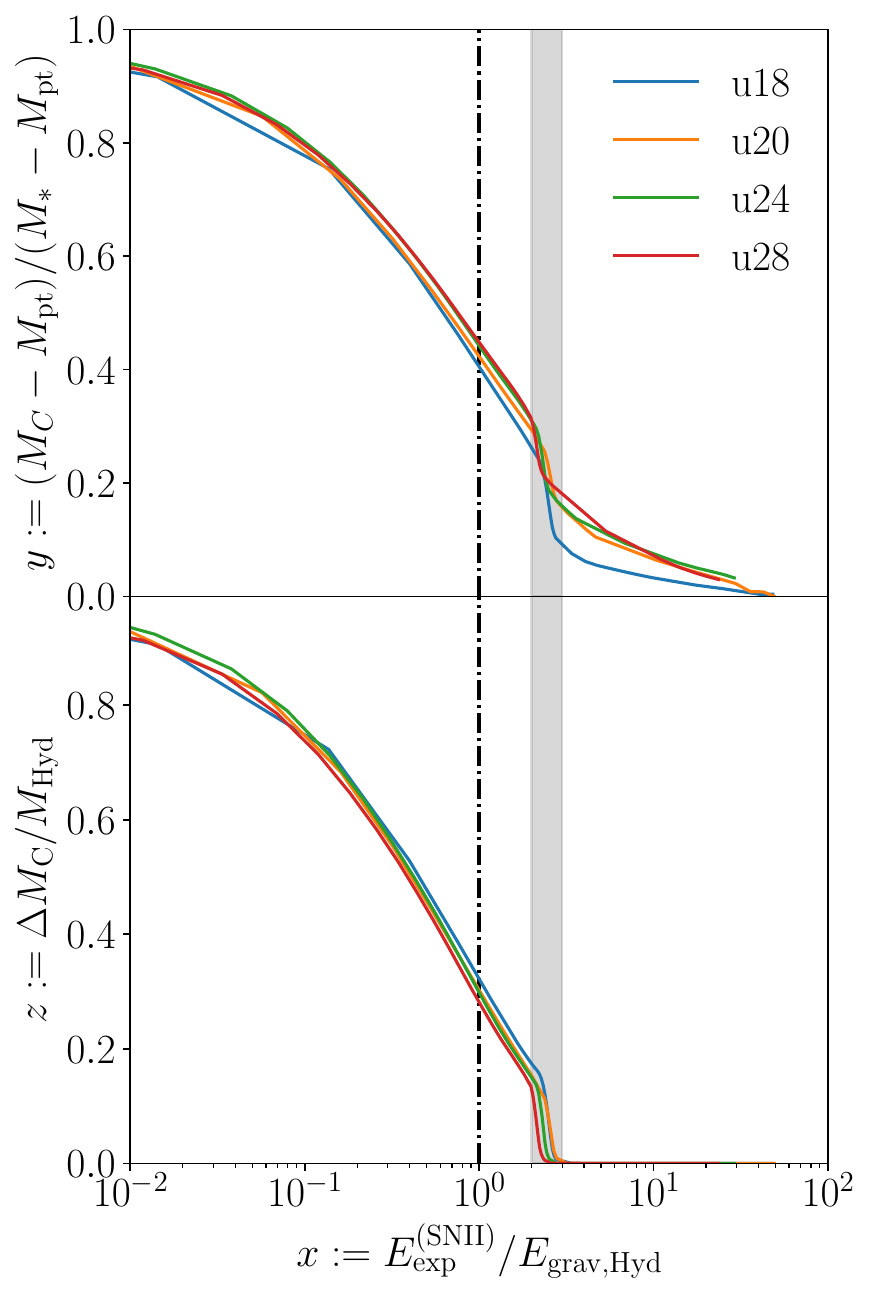}
    \caption{As in Figure \ref{fig:eachprog_Eex_MC_dM}, the axes are normalized as follows. The horizontal axis, x, is the explosion energy of the SNII progenitor divided by its hydrogen envelope’s binding energy, $E_\mathrm{grav,Hyd}$.  And the dashed line at x = 1 marks their equality. The vertical axis in the top panel is $y := (M_C - M_\mathrm{pt})/(M_* - M_\mathrm{pt})$, where $M_*$ is the total progenitor mass and $M_\mathrm{pt}$ is the point source mass (see \ref{subchap:softened_method}). In the bottom panel, the vertical axis is $z := \Delta M_\mathrm{C}/M_\mathrm{Hyd}$, with $M_\mathrm{Hyd}$ the hydrogen envelope mass and $\Delta M_\mathrm{C}$ defined in Figure \ref{fig:u18_Eex_MC_dM}.}
    \label{fig:eachprog_norm_Eex_MC_dM}
\end{figure}

As shown in Figure \ref{fig:eachprog_Eex_MC_dM}, all the models behave similarly.   
In order to elucidate the underlying common physics, we plot these curves but normalized with characteristic mass and energy scales (Figure \ref{fig:eachprog_norm_Eex_MC_dM}).
The horizontal axis is normalized by the binding energy of the hydrogen envelope of each progenitor $(x = E_{\rm exp} / E_{\rm grav,Hyd})$. In the top panel, we plot the fallback fraction defined as
\begin{align}
    y = \frac{M_C - M_\mathrm{pt}}{M_* - M_\mathrm{pt}},
    \label{eq:norm_MC}
\end{align}
where $M_*$ is the total mass of the progenitor before the explosion. Note that Equation (\ref{eq:norm_MC}) is the ratio of the total mass of progenitors excluding the first formed neutron star and the remnant mass after the explosion.
Whereas in the bottom panel we plot the excess fallback fraction defined as
\begin{align}
    z = \frac{\Delta M_{\rm C}}{M_{\rm Hyd}}.
\end{align}

By normalizing in the way described above, we can see that all the models almost coincide, and the {\it transition} region is concentrated in the range of $2\lesssim x \lesssim 3$. This range means that even if the progenitor is changed, the transition region exists where the explosion energy is at most $2-3$ times the binding energy of the hydrogen envelope. If the explosion energy exceeds this, all of the hydrogen envelope will be ejected.

In addition, it appears that the trajectories of all progenitors converge on the low-energy side ($x\lesssim 2$). The reason may be that the hydrogen envelope accounts for most of the total mass of the progenitor, and on the low-energy side, the mass of the hydrogen envelope dominates most of the remnant mass.
On the other hand, all the progenitors except u18 follow a similar trajectory on the high-energy side ($3\lesssim x$).

One reason why the u18 lies outside this region is the difference in the C/O core structure. The final column of Table \ref{tab:prog_val} represents the ratio of mass to radius for the C/O core. This corresponds to the compactness of the C/O core \citep[e.g.,][]{O'Connor2011ApJ}. The CO mass-to-radius ratio for u18 is significantly smaller than that of the other models.

The vertical axis of the lower panel is the normalized variable $z$, which is obtained by dividing $\Delta M_\mathrm{C}$ by the hydrogen envelope mass $M_\mathrm{Hyd}$ (Table \ref{tab:prog_val}) of each progenitor. This normalized variable $z$ is defined such that $z = 0$ when the hydrogen envelope does not fall to the centre, whereas $z = 1$ when it completely falls.
We find that $z$ takes a value of $0 - 0.15$ in the transition region ($2\lesssim x \lesssim 3$). 
Also, focusing on the low-energy side, we can see that even with an explosion energy of only $1\%$ of the binding energy of the hydrogen envelope, $10\%$ of the mass of the hydrogen envelope is ejected. This corresponds to the fact that the outer edge of the hydrogen envelope has a shallow gravitational potential, and even a tiny explosion of energy can unbind it.

\begin{figure}
    \centering
    \includegraphics[width=1.\linewidth]{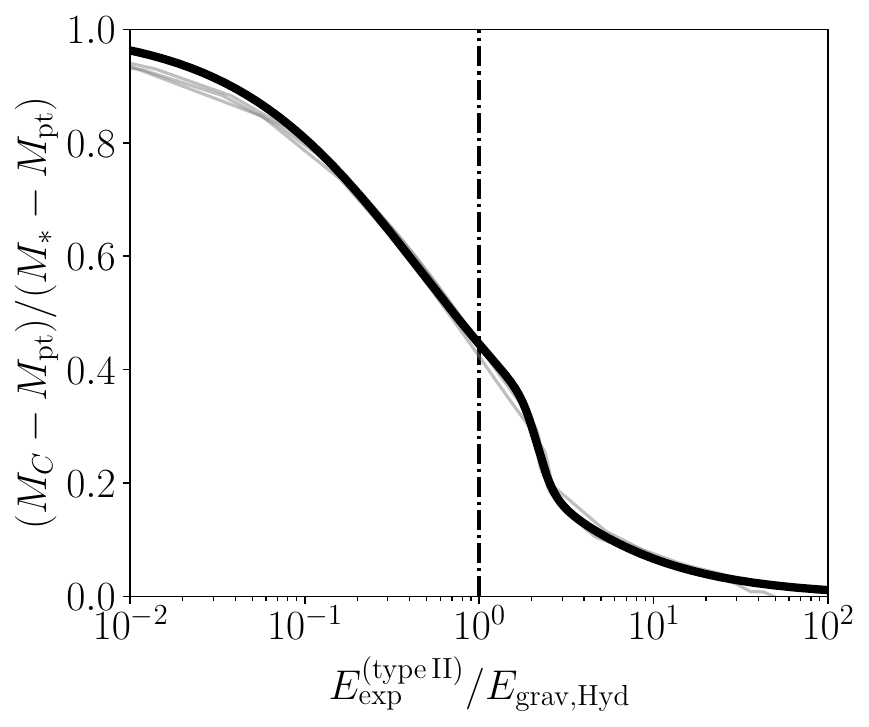}
    \caption{
    Same as the upper panel of Figure \ref{fig:eachprog_norm_Eex_MC_dM}, but for the results of u20, u24, u26, and u28. The grey lines show the results of each model, and the black line represents the fitting curve obtained using Equation (\ref{eq:fitting_function}) with parameters $w = 0.85$, $x_1 = 0.46$, $k_1 = 0.81$, $x_2 = 2.16$, and $k_2 = 9.42$.}
    \label{fig:fit_norm_Eex_Mc}
\end{figure}

As shown in Figure \ref{fig:fit_norm_Eex_Mc}, the remnant mass $M_C$ can be reasonably fitted via
\begin{align}
\label{eq:Mrem_from_Mstar_Eexp_EgHenv}
M_{C} = (M_{*} - M_{\rm pt}) f(x) + M_{\rm pt},
\end{align}
except for the u18 case.
Here, we employ the following fitting function for $f(x)$,
\begin{align}
\label{eq:fitting_function}
f(x) =
\frac{w}{1 + \left( \dfrac{x}{x_1} \right)^{k_1}}
+
\frac{1-w}{1 + \left( \dfrac{x}{x_2} \right)^{k_2}}.
\end{align}
Here,  $w=0.85,\,x_1=0.46, \,k_1=0.81, \,x_2=2.16, \, k_2=9.42$.

\section{Discussion and Conclusion}
\label{sec:discussion_and_conclusion}

It remains challenging to consistently link progenitors to final explosion energies and remnant masses, including fallback.
Two difficulties arise: (i) reproducing core collapse requires sophisticated simulations with neutrino transport and detailed microphysics, and (ii) collapse ($\sim 1 \,{\rm s}$) and fallback ($\sim 10^6 \, {\rm s}$) occur on vastly different timescales, making fully consistent calculations prohibitive.

By contrast, calculations of shock propagation through the hydrogen envelope are computationally cheaper, yet crucial for determining the remnant mass.
The reverse shock generated at the H/He interface can either escape or fall back, substantially altering the final compact-object mass.
Conventional fallback studies, however, imposed hollowed-out inner boundaries that spuriously reflect shocks and generate unphysical waves in 1D–3D, highlighting the need to revise boundary treatments.

In this work, we performed fallback simulations
with boundary conditions that avoid artificial reflections by replacing the inner core with a softened point mass surrounded by a tenuous atmosphere, avoiding artificial reflections.
We considered $Z=10^{-4}Z_\odot$ progenitors of $M_\mathrm{ZAMS}=18$–$28\,M_\odot$ \citep{Woosley2002RvMP}, with and without hydrogen envelopes, and drove explosions via a thermal bomb with variable injected energy.

Observationally, Type II SNe show a broad distribution of explosion energies, typically ranging from $10^{50}-10^{52}\,\mathrm{erg}$ with a median around $6\times10^{50}\,\mathrm{erg}$ \citep{Martinez2022A&A}.
Moreover, recent candidates for failed SNe \citep{Neustadt2021MNRAS,Beasor2024ApJ,De2024arXiv} have suggested the presence of even weaker explosions, possibly as low as $10^{47}$–$10^{49}\,\mathrm{erg}$.
However, no previous fallback calculations have consistently covered this full energy range, partly because such weak and strong explosions were beyond the range of exploration and partly due to numerical artifacts from reflected reverse shocks at the H/He interface.

In the present study, we focus not only on the effect of the C/O core but also on the binding energy of the hydrogen envelope, which plays a crucial role in regulating the fallback.
We identify a consistent transition region across all progenitor models: when the explosion energy reaches $\sim2$–$3$ times the envelope binding energy, the reverse shock begins to fall back and the remnant mass rises sharply.
We have quantified this behavior over a wide range of explosion energies ($10^{48}$–$10^{52},\mathrm{erg}$) and formulated a new expression for the remnant mass as a function of the explosion energy and the hydrogen-envelope binding energy.
This formulation provides a robust and continuous description of the remnant-mass growth through fallback, bridging low- and high-energy explosions in a unified framework.

These results are also broadly applicable to population studies.
Equation (\ref{eq:Mrem_from_Mstar_Eexp_EgHenv}) provides a simple and direct way to calculate the remnant mass from the progenitor mass, hydrogen-envelope binding energy, and explosion energy, unlike conventional prescriptions (e.g., \citealt{Fryer2012ApJ}) that assume a one-to-one relation between the C/O core mass and remnant properties.
This relation can be readily used in population-synthesis models to estimate the distribution of black holes and neutron stars from the statistical variations in explosion energy and progenitor mass.
It therefore offers a practical and physically motivated prescription for modeling the demographics of compact remnants across a wide energy range and metallicity environments.

Furthermore, the fallback mass determined by this model also affects the mass and composition of the ejecta, which is important for galactic chemical and dust evolution.
A larger fallback mass leads to less ejected metals, while more energetic explosions can eject heavy elements more efficiently.
By connecting explosion energy, fallback, and ejecta composition, this framework naturally links remnant formation with dust evolution in supernovae and young galaxies.

In summary, this study provides a consistent description of how hydrogen envelopes and explosion energy shape the final remnant mass across a wide energy range.
The results are directly applicable to stellar-population and dust-formation models and offer an important step toward connecting small-scale supernova physics with large-scale galactic evolution.

\section*{Acknowledgements}

SK thanks A. Tanikawa, H. Umeda, T. Shigeyama, H. Kawashimo, C. Qu, E. Hattori, and T. Tokuno, for fruitful discussion. This work is supported by JST SPRING, Grant Number JPMJSP2108. This work is also supported by JSPS KAKENHI grant Nos. 21K13964, 23K22534, 24H02245, 24K00668, and 25K01035.

\section*{Data Availability}
The data underlying this article will be shared on reasonable request to the corresponding author.



\bibliographystyle{mnras}
\bibliography{sn_fallback} 




\appendix
\section{Implementation of the Self-gravity Term}

\label{app:gravity}
In this section, we discuss how self-gravity is reproduced. The cell centre index is denoted as an integer and the cell surface index as a half-integer. First, it is essential to calculate the enclosed masses for the self-gravity calculation correctly,
\begin{align}
	M_{i+\frac{1}{2}} = \sum^{ie}_{is} \Delta M_i + M_{is - \frac{1}{2}}. 
\end{align}
Since it is 1D spherically symmetric, the first term is written as,
\begin{align*}
	\Delta M_i = \int_{V_i} 4\pi r^2\rho dr .
\end{align*}
The integral can be computed with quadratic precision by using trapezoidal integrals. For clarity, the subscripts are rewritten as $i+1/2 \rightarrow p, i-1/2\rightarrow m$. The trapezoidal integral is written explicitly as,
	\begin{align}
		\Delta M_i & = \int^{r_p}_{r_m} 4\pi r^2 \rho(r) dr        \nonumber\\
		           & \simeq 2\pi(r_p^2\rho_p+r_m^2\rho_m)(r_p-r_m) ,
	\end{align}
	where $\rho_p, \rho_m$ are left density and right density,
\begin{align}
\rho_{m} &= \rho_{i-1} + \frac{\rho_i - \rho_{i-1}}{r_i - r_{i-1}} (r_{i-\frac{1}{2}} - r_{i-1}),\\
\rho_{p} &= \rho_{i} + \frac{\rho_{i+1} - \rho_{i}}{r_{i+1} - r_{i}} (r_{i+\frac{1}{2}} - r_{i}).
\end{align}
In the case of spherically symmetric, self-gravity is
	\begin{align}
		g_i = - \frac{GM_i}{r^2}. 
	\end{align}
So, it is possible to calculate the gravity if the enclosed mass $M_i$ is obtained. 
Next, we describe how the gravity term is incorporated in {\tt Athena++}.
The gravity term is involved in the momentum and energy conservation equations. Specifically,
\begin{align}
\label{mom_grav}
\pdv{\rho \bm{u}}{t} + \nabla\cdot[\rho\bm{u}\bm{u}+\bm{P}] &= \rho \bm{g},\\
\label{ene_grav}
\pdv{e_*}{t} + \nabla \cdot [(e_* + p)\bm{u}] &= \rho \bm{u}\cdot\bm{g},
\end{align}
where $e_* = k + u = \rho \bm{u}^2 + p / (\gamma - 1)$. Now, the physical quantity $\rho,e_*,\rho \bm{u}$ in the code is defined at the cell centre. The position of this cell centre is determined by
	\begin{align}
		r_{cc} = \frac{\int r dV_i}{\int dV_i}.
	\end{align}
 Therefore, the value of the right-hand side in (\ref{mom_grav}) and (\ref{ene_grav}) must be {the gravity term at the centre of gravity, considering the cell gradient}. The term that takes this effect into account is enclosed by $\left\langle \right\rangle$. 

Below, we put  
\begin{align}
\dot{m}_{gi} &= \left \langle \rho \bm{g} \right \rangle_i = \frac{\int \rho \bm{g} r^2dr}{\int r^2dr}, \\
\dot{e}_{gi} &= \left \langle \rho \bm{u}\cdot\bm{g} \right \rangle_i = \frac{\int \rho \bm{u}\cdot\bm{g}r^2dr}{\int r^2dr}. 
\end{align}

In this calculation, $e^{n+1}_i, (\rho u)^{n+1}_i $ at $n+1$ cycle is calculated using $e^{n}_i, (\rho u)^{n}_i, \dot{e}^n_{gi}, \dot{m}^n_{gi}$ at $n$ cycle,
\begin{align}
e^{n+1}_i &= e^{n}_i + \dot{e}^n_{gi}dt ,\\
(\rho u)^{n+1}_i &= (\rho u)^{n}_i + \dot{m}^n_{gi}dt.
\end{align}

\label{lastpage}
\end{document}